\documentclass[12pt,a4paper]{article}
\usepackage{amssymb}
\usepackage{graphicx}

\newcommand{\mpi}{M_{\pi}}

\newcommand{\Mpi}{M_{\pi}}
\newcommand{\Fpi}{F_{\pi}}

\newcommand{\til}{\tilde}

\newcommand{\la}{\lambda}

\newcommand{\be}{\begin{equation}}
\newcommand{\ee}{\end{equation}}
\newcommand{\bea}{\begin{eqnarray}}
\newcommand{\eea}{\end{eqnarray}}
\newcommand{\bdm}{\begin{displaymath}}
\newcommand{\edm}{\end{displaymath}}

\newcommand{\mr}{\mathrm}

\newcommand{\lb}{\bar\ell}

\def\fs{\; \; .}
\def\co{\; \; ,}

\begin{document}
\renewcommand{\today}{April 6, 2004}

\title{\Large\bf An asymptotic formula for the pion decay constant in a
  large volume} 
\author{ 
 Gilberto Colangelo and Christoph Haefeli \vspace{1cm}\\
{\small
Institut f{\"u}r Theoretische Physik, Universit{\"a}t Bern}
\\
{\small Sidlerstr. 5, 3012 Bern,  Switzerland } }
\maketitle
\begin{abstract}
We derive an asymptotic formula {\it \`a la} L\"uscher for the finite
volume correction to the pion decay constant: this is expressed as an
integral over the $\langle 3 \pi | A_\mu|0 \rangle$ amplitude after proper
subtraction of the pion pole contribution. We analyze the formula
numerically at leading and next-to-leading order in the chiral expansion.
\end{abstract}
\thispagestyle{empty}
\setcounter{page}{0}

\paragraph{1. Introduction}
The analytical study of finite volume effects is becoming of increasing
importance as lattice calculations with dynamical fermions approach smaller
quark masses and aim at higher precision. Since these effects are dominated
by the lightest particles in the spectrum, the pions, and by their long
distance dynamics, one can study them in the framework of chiral
perturbation theory (CHPT) \cite{Gasser:1987zq}. A number of analyses of
these effects in different quantities have recently appeared in the
literature \cite{Colangelo:2003hf,recent}. One of these concerned the case
of the pion mass \cite{Colangelo:2003hf} and has shown that a leading order
calculation may receive very large corrections from the next-to-leading
contribution even for small values of the quark masses, whereas even higher
order corrections behave according to expectations and show a convergent
behaviour. This accurate study of the convergence of the chiral series has
been made possible by the use of L\"uscher's asymptotic formula for the
pion mass \cite{Luscher:1985dn}. The formula relates its leading
finite-volume corrections to an integral over the $\pi \pi$ scattering
amplitude in infinite volume. Since the latter is known to
next-to-next-to-leading order in the chiral expansion \cite{BCEGS}, it is
straightforward to evaluate L\"uscher's formula to the same order in the
chiral expansion.

In view of the results for the pion mass, the question arises if one can
derive similar asymptotic formulae also for other quantities: as we will
show in what follows, this is the case. In the present article we
concentrate on $F_\pi$, derive an asymptotic formula which relates
it to the infinite-volume $\langle 3 \pi | A_\mu|0 \rangle$ amplitude and
analyze it numerically using the next-to-leading order calculation of this
amplitude \cite{Colangelo:1996hs}. The results show again large
next-to-leading order corrections -- in this case we cannot explore the
chiral expansion further because the two-loop calculation of the 
$\langle 3 \pi | A_\mu|0 \rangle$ amplitude is not yet available.

\paragraph{2. The asymptotic formula}
Denote by $F_{\pi,L}$ the pion decay constant in a box of size $L$. 
The asymptotic formula for $\Delta F_\pi = F_{\pi,L}-F_\pi$ can then be 
written as:
\be
\Delta F_\pi = \frac{3}{8 \pi^2 M_\pi L} \int_{-\infty}^{\infty}d y
\; e^{-\sqrt{\mpi^2+y^2} L} N_F(iy) +O(e^{-\bar ML}) \co
\label{eq:Fpi}
\ee
where $\bar{M}\!\geq\!\sqrt{3/2}\:\Mpi$ and the amplitude $N_F(\nu)$ is
defined as follows. Consider the amplitude for creation of three pions out
of the vacuum with the axial current: 
\bea
\langle \pi^1(p_1) \pi^1(p_2) \pi^3(p_3) | A^3_\mu(0) | 0 \rangle &=&
(p_1+p_2)_\mu G(s_1,s_2,s_3) \label{eq:A3pi}\\
&+&(p_1-p_2)_\mu H(s_1,s_2,s_3)+p_{3 \, \mu} F(s_1,s_2,s_3) \nonumber \co
\eea
where the superscripts on the pion states and axial current are isospin
indices and $G$, $H$ and $F$ are three scalar amplitudes of the variables
$s_1$,$s_2$ and $s_3$, with $s_1=(p_2+p_3)^2$ and cyclic permutations
\cite{Colangelo:1996hs}. From the amplitude (\ref{eq:A3pi}) one can
construct the combination which has two of the outcoming pions in an
$I=0$ state (the explicit relation is given below)
\bea
\langle (2 \pi )_{I=0} \pi^3(p_3) | A^3_\mu(0) | 0 \rangle &=&
(p_1+p_2)_\mu G_0(s_1,s_2,s_3)  \\
&+&(p_1-p_2)_\mu H_0(s_1,s_2,s_3)+p_{3 \, \mu} F_0(s_1,s_2,s_3) \nonumber
\fs 
\eea
This amplitude contains a pole in the unphysical region, for
$(p_1+p_2+p_3)^2=Q^2=M_\pi^2$, which needs to be removed before proceeding
further. We define
\bea
\langle (2 \pi )_{I=0} \pi^3(p_3) | A^3_\mu(0) | 0 \rangle_S &=&
\langle (2 \pi )_{I=0} \pi^3(p_3) | A^3_\mu(0) | 0 \rangle \nonumber \\
&-&Q_\mu \frac{i F_\pi T^{I=0}(s_3,s_1-s_2)}{M_\pi^2-Q^2}
\label{eq:subtr} \co
\eea
where $T^{I=0}(s,t-u)$ is the $\pi \pi$ scattering amplitude with isospin
zero in the $s$ channel. We need the subtracted amplitude in the
forward kinematic configuration, i.e. for $p_1=-p_2$, $s_3=0$, where it
becomes a function of one variable only, $\nu=(s_2-s_1)/(4 \mpi)$:
\be
p_3^\mu \langle (2 \pi )_{I=0} \pi^3(p_3) | A^3_\mu(0) | 0
\rangle_{S\; |_{p_1=-p_2}} = 2  \mpi \nu h_0(\nu) + \mpi^2 \bar f_0(\nu)
\co
\label{eq:A3pinu}
\ee
where
\[
h_0(\nu)=H_0(2 \mpi(\mpi \!- \! \nu),2 \mpi(\mpi \!+\! \nu),0)
\]
and analogously for $\bar f_0$ and where the bar on the $F_0$ form factor
denotes that it is defined after subtraction of the pion pole (the form
factor $H_0$ remains unaffected by the subtraction). The amplitude $N_F$ which
enters the asymptotic formula for the finite volume corrections to $F_\pi$
is defined as
\be
N_F(\nu)=-i \left( 2 \nu h_0(\nu) + \mpi \bar f_0(\nu) \right) \fs
\label{eq:NF}
\ee
The amplitudes $H_0$ and $F_0$ can be expressed in terms of $F$, $G$
and $H$ appearing in (\ref{eq:A3pi}):
\bea
F_0(s_1,s_2,s_3)&\!\!=\!\!& 3F_{123}+G_{231}+G_{312}-H_{231}+H_{312} \co
\nonumber\\
H_0(s_1,s_2,s_3)&\!\!=\!\!& 3H_{123}+\frac{1}{2}
\left[F_{231}-F_{312}-G_{231}+G_{312}-H_{231}-H_{312} \right]  , \; \;
\eea
where $X_{ijk}=X(s_i,s_j,s_k)$ with $X=F,G,H$.

\paragraph{3. Outline of the derivation}
The derivation of this formula is in large parts analogous to the
derivation of the formula for the pion mass, due to L\"uscher
\cite{Luscher:1985dn}. In the following we simply outline the necessary
steps to prove the formula and refer the reader to the paper of
L\"uscher for details.  The starting point of the analysis is that one can
rely on an effective Lagrangian description of the relevant physics and
analyze these finite volume effects in CHPT. As observed by L\"uscher, the
precise form of the effective Lagrangian is never needed in the proof -- on
the other hand, it is very useful to have it available if one wants to
understand in concrete terms these effects. As was shown by Gasser and
Leutwyler one can rigorously derive the consequence of chiral symmetry also
if the system is closed inside a large finite volume with the help of the
effective Lagrangian technique \cite{Gasser:1987zq}. In particular the form
of the local effective Lagrangian remains unchanged, and the only
difference with respect to infinite volume calculations comes from the
propagator for the pion field which becomes periodic in all spatial 
directions 
\be
G(x^0,\vec{x})=\sum_{\vec{n} \in \mathbb{Z}^3}G_0(x^0,\vec{x}+\vec{n}L)
\label{eq:G}
\ee
where $G_0(x)$ is the propagator in infinite volume.

The first step in L\"uscher's proof of the asymptotic formula for the pion
mass consists in showing that, for a generic loop diagram contributing to
the self energy of the pion, the dominating finite volume effect is
obtained if one takes all propagators in infinite volume ($G(x) \to
G_0(x)$) except one, for which only the terms with $|\vec{n}|=1$ in the sum
in (\ref{eq:G}) should be kept\footnote{More precisely: this concerns only
  propagators which are contained in at least one loop,
  cf. \cite{Luscher:1985dn} }. The sum of all possible contributions of 
this form from all possible loop diagrams gives the leading finite volume
corrections to the pion mass. The same conclusion is valid also for the
Feynman diagrams which renormalize the coupling between the axial current
and the pion -- the fact that in this case one of the external legs is the
axial current instead of a pion does not touch the argument at all.

\begin{figure}[t]
\includegraphics[width=6.5cm]{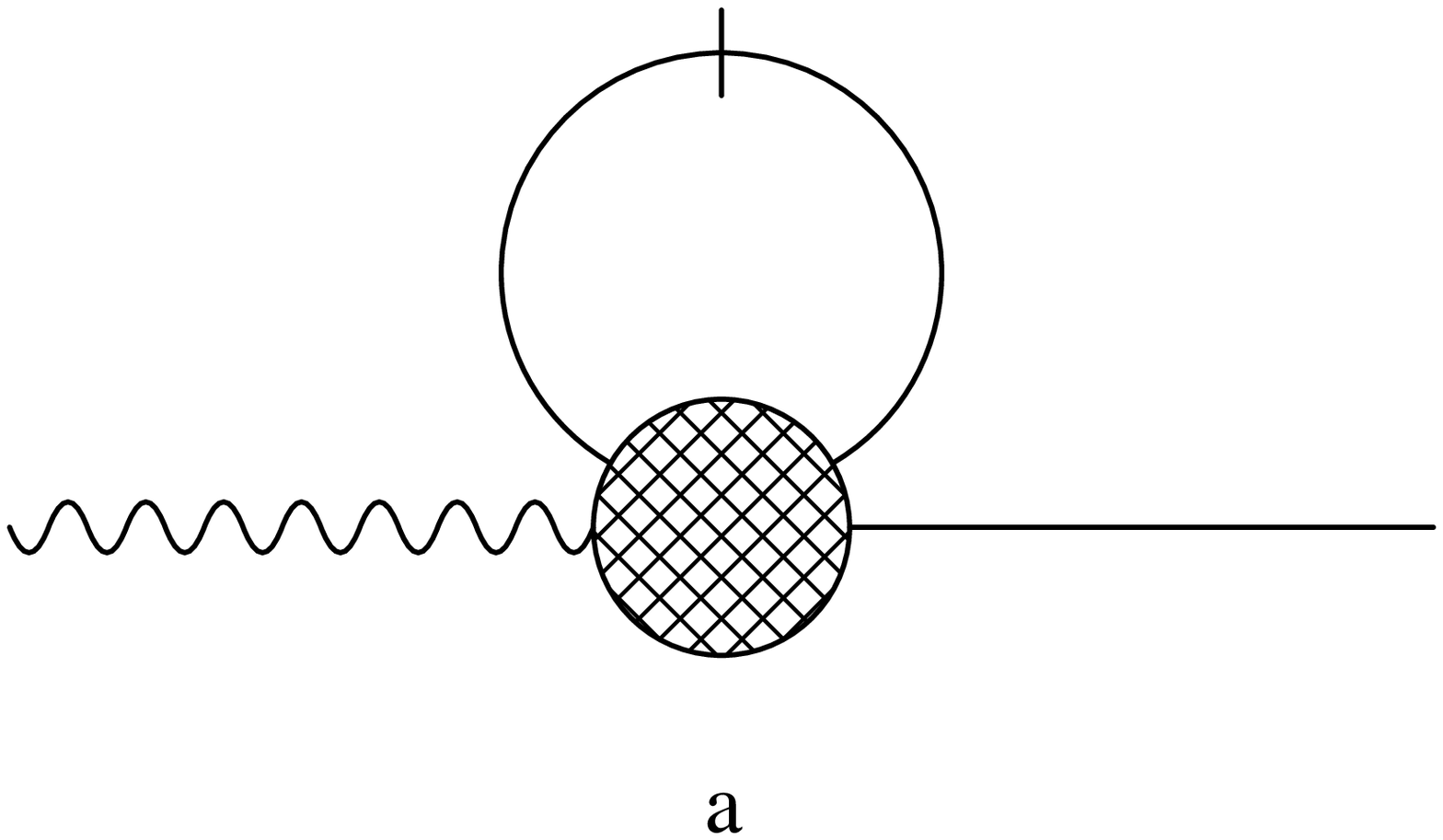} \hfill
\includegraphics[width=6.5cm]{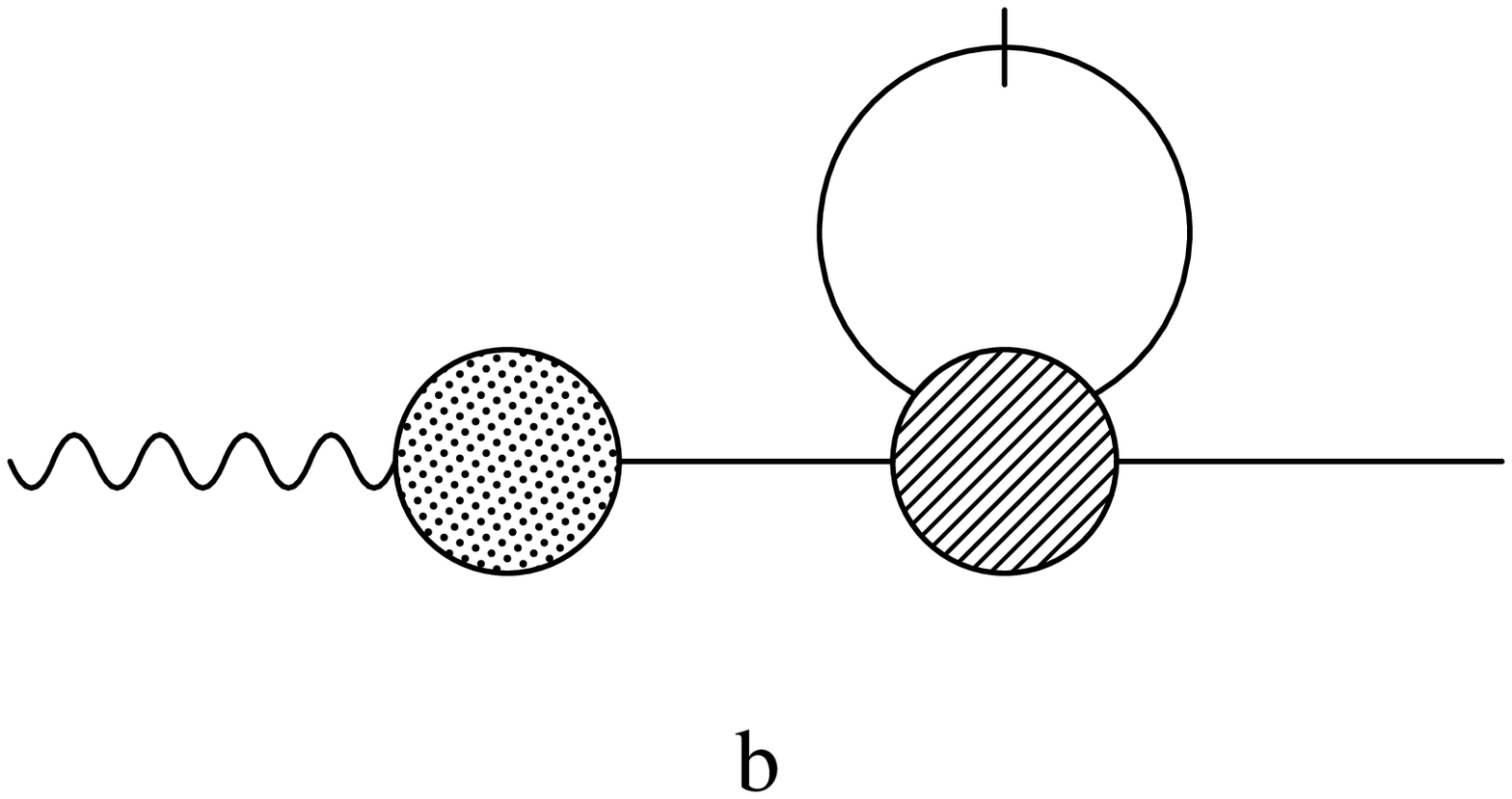}
\caption{\label{fig:FDS}
Graphical representation of the asymptotic formula. The wiggly (straight)
line represents the axial current (pion). The dash on the
propagator means that it is taken in finite volume (only the contribution
with $|\vec{n}|=1$ in the sum (\ref{eq:G})). Diagram a (b) illustrates the
correction to $F_\pi$ (the shift of the pole position) due to finite
volume.} 
\end{figure}
The second step in the proof consists in showing, by modifying the
integration contour in the complex plane, that this leading contribution
can be written in a very compact form, as an integral over an amplitude
(the $\pi \pi$ scattering amplitude in the case of the pion mass) defined
in Minkowski space, analytically continued to complex values of its
arguments. Again, the same argument applies also to the case of the pion
decay constant: in this case, in all possible loop graphs that renormalize
the pion coupling to the axial current we have to single out one internal
pion propagator, break it up and put the resulting two external legs on
shell. The relevant amplitude in this case is the $\langle 3 \pi | A_\mu|0
\rangle$ amplitude, as illustrated in Fig.~\ref{fig:FDS}a -- the weight
function which appears in the integral is however exactly the same as in
the pion mass case.

The kinematic configuration in which the amplitude must be evaluated is
also the same and corresponds, for the $\pi \pi$ amplitude, to forward
scattering. The $\langle 3 \pi | A_\mu|0 \rangle$ amplitude is however
singular for this kinematics because of a pole due to one-pion exchange
among the axial current and the three outgoing pions. This singularity does
not belong to the finite volume corrections to $F_\pi$ and should be
subtracted. The reason for the presence of this pole can be explained
as follows: the $\langle \pi|A_\mu|0\rangle$ amplitude is defined as the
residue at the pion pole of a two-point function of the axial current and
any interpolating field for the pion:
\bea
\langle \pi^a(q) |A^b_\mu|0\rangle &=& \lim_{q^2 \to \mpi^2} (M_\pi^2-q^2)
i q_\mu \delta^{ab} P(q)
\\
P(q) &=& N_{\phi} q^\mu
\int dx e^{iqx} \langle0 | T \phi^1_\pi(x) A^1_\mu(0) |0 \rangle \nonumber \co
\eea
with $N_\phi$ the proper normalization factor which depends on the field
$\phi_\pi$. In finite volume both the residue as well as the position of
the pole are shifted. Ignoring the latter shift corresponds to multiplying
$P_L(q)$ by $(M_\pi^2-q^2)$ and not by the correct $(M_{\pi,L}^2-q^2)$ and
then taking the limit $q^2 \to \mpi^2$. The result, expanded to the leading
term for asymptotically large volumes, contains a pole for $q^2=\mpi^2$
\be
(M_\pi^2-q^2) P_L(q) \sim (M_\pi^2-q^2) \frac{F_{\pi,L}}{M_{\pi,L}^2-q^2}
= F_{\pi,L} - \frac{F_\pi \Delta M_\pi^2}{M_{\pi}^2-q^2} + \ldots \co
\ee
where $\Delta \mpi^2= M_{\pi,L}^2 -\mpi^2$ is also evaluated to leading
order. Since the shift in the pion mass is known and given by L\"uscher's
formula, we can subtract the pole (which is illustrated in
Fig.~\ref{fig:FDS}b) and get the correct finite-volume value of the pion
decay constant. The result leads to the subtraction prescription given in
the previous section.

\paragraph{4. The coupling constant $G_\pi$}
The formula presented here for $F_\pi$ can be extended with obvious
modifications also to other quantities, e.g. like $G_\pi$, the coupling
constant of the pion to the pseudoscalar quark bilinear 
$P^i= \bar q i \gamma_5 \tau^i q$
\be
\langle 0 |P^i(0) | \pi^k \rangle = \delta^{ik} G_\pi \fs
\ee
In this case the amplitude that should replace $N_F(\nu)$ in the analogue
of Eq.~(\ref{eq:Fpi}) is the subtracted $P \to 3 \pi$ amplitude in the
limit $p_1=-p_2$:
\be
N_G(\nu)= \lim_{p_1\rightarrow-p_2} \left[\langle (2 \pi )_{I=0} \pi^3(p_3) 
| P^3 (0) | 0
\rangle  - \frac{ G_\pi T^{I=0}(s_3,s_1-s_2)}{M_\pi^2-Q^2} \right] \fs
\ee
In this particular case the Ward identity ($\hat m \equiv (m_u+m_d)/2$)
\be
F_\pi M_\pi^2 = \hat m G_\pi \co
\label{eq:WI}
\ee
which also holds in finite volume, makes the use of such a formula
unnecessary: from the finite-volume version of Eq.~(\ref{eq:WI}) one
immediately obtains 
\be
\frac{\Delta G_\pi}{G_\pi} =:R_G= R_F+2 R_M \co
\label{eq:DeltaWI}
\ee
where $R_M$ is the relative shift for $M_\pi$. On the other hand, since we
have an explicit expression for all three relative shifts for large
volumes, Eq.~(\ref{eq:DeltaWI}) can be used as a nontrivial check on the
asymptotic formulae. Indeed, all three relative shifts can be expressed as
an integral with the same weight function, and Eq.~(\ref{eq:DeltaWI}) can
be satisfied only if the same relation holds among the integrands:
\be
N_G(\nu)=N_F(\nu)-\frac{F_\pi}{M_\pi} F(\nu) \co
\ee
where $F(\nu)=T^{I=0}(0,\nu)$ is the forward scattering amplitude appearing
in L\"uscher's formula for $M_\pi$. It is easy to verify that this relation
follows from the Ward identity\footnote{Notice that in the definition of
  $N_F$, Eqs.~(\ref{eq:A3pinu},\ref{eq:NF}), the $ \langle 3
  \pi|A_\mu|0\rangle $ amplitude is multiplied with $p_3^\mu$ and not with
  $Q^\mu$ as in this Ward identity.} 
\be
-i Q^\mu \langle \pi^1(p_1) \pi^1(p_2) \pi^3(p_3) | A^3_\mu(0) | 0 \rangle= 
\hat m \langle \pi^1(p_1) \pi^1(p_2) \pi^3(p_3) | P^3(0) | 0 \rangle \co
\ee
once the limit to the relevant kinematical
configuration is taken and if one properly accounts for the pole at
$Q^2=M_\pi^2$ present in both amplitudes.

\paragraph{5. The asymptotic formula in chiral perturbation theory}
As was shown in \cite{Colangelo:2003hf}, the L\"uscher formula for the pion
mass can be used very conveniently in combination with the chiral expansion
for the $\pi \pi$ scattering amplitude. The same can be done for $F_\pi$
using the chiral expansion for the infinite-volume $\langle 3 \pi | A_\mu|0
\rangle$ amplitude, which has been calculated up to next-to-leading order
in \cite{Colangelo:1996hs}. The chiral expansion for the amplitude $N_F$
reads
\be
N_F(\nu) = \frac{\mpi}{F_\pi} \left[ N_2^F(\tilde \nu) + \xi N_4^F(\tilde \nu) 
  +O(\xi^2) \right] \co
\ee
where $\xi=(\mpi/4\pi F_\pi)^2$ and $\tilde \nu = \nu/\mpi$, and translates
into a corresponding expansion for $\Delta F_\pi$ 
\be
R_F:= \frac{ \Delta F_\pi}{F_\pi}= \frac{6}{\lambda} \left[ \xi
  I_2^F(\lambda) + \xi^2 I_4^F(\lambda) +O(\xi^3) \right] \co
\label{eq:RF}
\ee
where $\lambda=\mpi L$.
The integrals $I_n$ can be given analytically in terms of a few basic
integrals:
\bea
I_2^F(\lambda)&=&-2 B^0(\lambda) \\
I_4^F(\lambda)&=&\left( 2 \lb_1 + \frac{4}{3} \lb_2 - 3 \lb_4 - \frac{7}{9}
\right) B^0(\lambda) 
+\left(-\frac{8}{3} \lb_1 -\frac{32}{3} \lb_2+ \frac{112}{9} \right)
B^2(\lambda) \nonumber \\ 
&+& \! \! \frac{4}{3}\left( R_0^0(\lambda) - R_0^1(\lambda) - 10 R_0^2(\lambda) 
 \right) 
- \frac{13}{6} R_0^{0 \prime}(\lambda)+\frac{8}{3} 
R_0^{1 \prime}(\lambda) + \frac{20}{3} R_0^{2 \prime} (\lambda) \, , \nonumber
\eea
where the integrals $B^{2k}$ and $R_i^k$ are defined as 
\be
B^{2k}(\la) = \int_{-\infty}^{\infty}d\til y\;\til y^{2k}\,
e^{-\sqrt{1+\til y^2}\la} =
{\Gamma(k+1/2) \over \Gamma(3/2) } \left( {2 \over \la} \right)^k
K_{k+1}(\la) \co
\ee
and
\be
R_0^{k (\prime)} (\la) = \left\{{\mr{Re}\atop\mr{Im}}\right.
\int_{-\infty}^{\infty} \hskip -0.5 cm d\til y\;\til y^k\,e^{-\sqrt{1+\til
    y^2}\,\la}\, g^{(\prime)} \left(2(1+{\rm i}\til y)\right)
\quad\mr{for}\left\{{k\;\mr{even}\atop k\;\mr{odd}}\right. \co
\label{eq:R0k}
\ee
with\footnote{The function $g(x)$ is related to the standard $\bar J$
  one-loop function through $g(x)= 16 \pi^2 \bar J(x \mpi^2)$.} 
\be
g (x)=\sigma \log \frac{ \sigma-1}{\sigma+1 }+2 \co \qquad
g^{\prime}(x)= \frac{1}{x} \left[
  \frac{2}{\sigma x} \log \frac{ \sigma-1}{\sigma+1 }-1 \right] \co
\ee
with $\sigma=\sqrt{1-4/x}$.
These integrals (with the only exception of the primed $R_0^{k}$) have
already been introduced in \cite{Colangelo:2003hf}.

\begin{figure}[t]
\includegraphics[width=12.9cm]{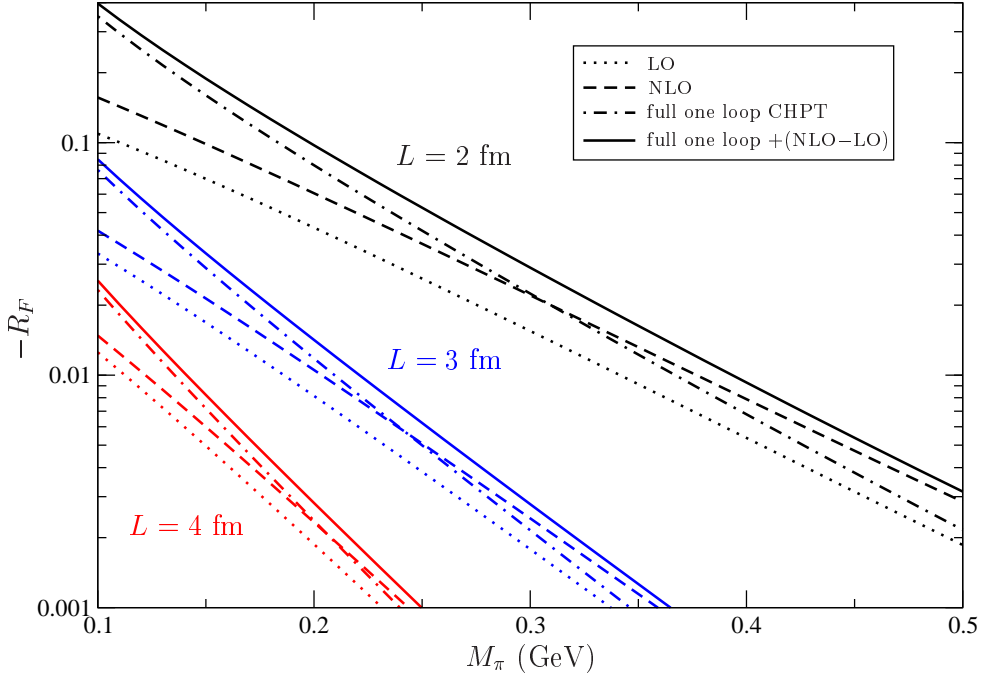}
\caption{\label{fig:fpi_mpi} The absolute value of the relative finite
volume correction $R_F=F_{\pi,L}/F_\pi-1$ as a function of $\mpi$ for
different volume sizes. We plot the leading (LO) and next-to-leading order
(NLO) in the chiral expansion of the asymptotic formula (\ref{eq:RF}) and
also the full one-loop result in CHPT (\ref{eq:Fpi1l}). The solid lines
show the sum of the full one-loop result and the NLO correction in the
asymptotic formula.} 
\end{figure}

We have evaluated numerically these corrections using the following values
for the chiral low energy constants \cite{CGL}:
\be
\lb_1=-0.4 \pm 0.6, \qquad \lb_2=4.3 \pm 0.1, \qquad 
\lb_4=4.4 \pm0.2 \fs
\label{eq:li}
\ee
The results are displayed in Fig.~\ref{fig:fpi_mpi} where we plot the
modulus of $R_F$ as a function of $\mpi$ for volume sizes between 2 and 4
fm. We have studied the uncertainties in $R_F$ which arise from the low
energy constants (\ref{eq:li}) and found that they are barely visible on
the plot -- we therefore omit them (in size they are similar to the
thickness of the lines).  In the figure we compare the evaluation of the
asymptotic formula to leading and next-to-leading order also to the full
one-loop calculation of Gasser and Leutwyler \cite{GaLeFSE1}, which can be
given in a very compact form:
\be
F_{\pi,L}=\;\Fpi\left[1-\; \xi\, \tilde g_1(\lambda)+O(\xi^2)\right]
\label{eq:Fpi1l}
\ee
where 
\be
\tilde g_1(\lambda) = \sum\nolimits' \int_0^\infty dx \;
e^{-{ 1 \over x}-{x\over 4}(n_1^2+n_2^2+n_3^2)\la^2} \co
\label{gtil}
\ee
where the prime indicates that the sum runs over all integer values of
$n_i$, excluding the term with all $n_i=0$.

In comparison to the pion mass, the finite volume corrections in
Eq.~(\ref{eq:Fpi1l}) are a factor 4 larger but negative -- the sign
difference is in accordance with the observation that in finite volume
chiral symmetry is restored, i.e. the pion becomes heavier and its decay
constant tends to vanish. Apart from this quantitative difference,
the numerical analysis gives results which are qualitatively similar to those
obtained for the pion mass \cite{Colangelo:2003hf}:
\begin{enumerate}
\item
the finite volume corrections are exponentially suppressed for large values
of $\mpi L$ and become negligible rather quickly;
\item 
the leading term in the chiral expansion of the asymptotic formula
receives large corrections even for the physical values of the quark masses
-- the similarity to the pion mass results makes us however think that 
the series will start to show a convergent behaviour at NNLO;
\item
the leading term in the asymptotic expansion also receives large
corrections from the subleading ones whenever the finite volume effects are
nonnegligible; 
\item
since subleading terms are important both in the chiral as well as in the
asymptotic expansion, the best estimate of the size of these finite-volume
corrections is obtained by {\it summing } the subleading effects in both
expansions, as shown by the solid curves in Fig.~\ref{fig:fpi_mpi}.
\end{enumerate}
For example, in a recent calculation of $F_\pi$ on the lattice~\cite{MILC}
with dynamical fermions a volume of $L=2.5$ fm size has been used, and pion
masses as low as $0.24$ GeV. For these values the finite volume corrections
evaluated with the asymptotic formula to NLO (LO) are 1.5\% (1.1\%),
whereas the full one-loop calculation gives 1.6\%. Adding both types of
subleading effects we find a total correction of 2\%. In Ref.~\cite{UKQCD},
$L=1.5$ fm and $M_\pi=0.4$ GeV were used: in this case the full one-loop
calculation gives a 3.4\% effect, whereas adding the NLO chiral corrections
we get to 4.5\%. For the parameters used in \cite{qq+q} finite-volume
effects are negligible.

\paragraph{6. Conclusions}
We have derived an asymptotic formula for the pion decay constant in a
finite large volume along the same lines as L\"uscher's formula for the
pion mass \cite{Luscher:1985dn}. The advantage offered by such a formula is
a relatively easy access to a study of higher order chiral corrections in
finite volume effects. We have evaluated these numerically and have shown
that in $F_\pi$ these corrections are large, analogously to what has been
found for $\mpi$ \cite{Colangelo:2003hf}. In the present case we could use
existing calculations of the relevant infinite-volume amplitude to evaluate
next-to-leading chiral corrections. Going one order higher in this
expansion would require the calculation of the $\langle 3 \pi | A_\mu| 0
\rangle$ amplitude to two loops in CHPT.

The asymptotic formula derived here immediately applies (after the
necessary but obvious modifications) to other similar quantities, like
$G_\pi$. As we have explicitly verified, the asymptotic formulae for
$F_\pi$ and $G_\pi$ satisfy a Ward identity that relates their ratio to
$M_\pi^2/\hat m$: if one extracts the finite-volume expression for $M_\pi$
from this Ward identity one recovers L\"uscher's formula. The formula
applies also to the decay constants of heavier mesons, like $F_K$. In the
latter case the study of these finite volume effects \cite{CH} is of direct
phenomenological interest in view of the recent application of the lattice
calculation of the $F_K/F_\pi$ ratio to the extraction of $V_{us}$
\cite{Marciano:2004uf} -- it is worth mentioning that for this application
the required precision of the lattice result is at the percent level. The
same formula can also be applied to the decay constants of yet
heavier mesons, like $f_D$ or $f_B$. In this case, however, the advantage
provided by the asymptotic formula with respect to a plain one-loop
calculation (as recently performed in~\cite{Arndt:2004bg}) will be of practical
relevance only if the knowledge of the low energy constants of the chiral
Lagrangian describing the coupling of heavy mesons to pions
\cite{heavy-light} is extended beyond leading order.

\paragraph{Acknowledgments}
We thank Stephan D\"urr, Heiri Leutwyler, Martin L\"uscher and Rainer
Sommer for useful discussions and/or comments on the manuscript. This work
is supported by the Swiss National Science Foundation and in part by RTN,
BBW-Contract No. 01.0357 and EC-Contract HPRN--CT2002--00311 (EURIDICE).


\begin{thebibliography}{99}
\bibitem{Gasser:1987zq}
J.~Gasser and H.~Leutwyler,
Nucl.\ Phys.\ B {\bf 307} (1988) 763.


\bibitem{Colangelo:2003hf}
G.~Colangelo and S.~Durr,
Eur.\ Phys.\ J.\ C {\bf 33} (2004) 543
[arXiv:hep-lat/0311023].

\bibitem{recent}
D.~Becirevic and G.~Villadoro,
Phys.\ Rev.\ D {\bf 69} (2004) 054010
[arXiv:hep-lat/0311028];\\
A.~Ali Khan {\it et al.}  [QCDSF-UKQCD Coll.],
arXiv:hep-lat/0312030;\\
M.~Guagnelli {\it et al.},
arXiv:hep-lat/0403009;\\
D.~Arndt and C.~J.~D.~Lin,
arXiv:hep-lat/0403012;\\
S.~R.~Beane,
arXiv:hep-lat/0403015.

\bibitem{Luscher:1985dn}
M.~L\"uscher,
Commun.\ Math.\ Phys.\ {\bf 104}, 177 (1986).

\bibitem{BCEGS}
J.~Bijnens {\it et al.},
Phys.\ Lett.\ B {\bf 374} 210 (1996)
[arXiv:hep-ph/9511397],\\
Nucl.\ Phys.\ B {\bf 508} 263 (1997)
[Erratum-ibid.\ B {\bf 517} 639 (1998)]
[arXiv:hep-ph/9707291].

\bibitem{Colangelo:1996hs}
G.~Colangelo, M.~Finkemeier and R.~Urech,
Phys.\ Rev.\ D {\bf 54} (1996) 4403
[arXiv:hep-ph/9604279].

\bibitem{CGL}
G.~Colangelo, J.~Gasser and H.~Leutwyler,
Nucl.\ Phys.\ B {\bf 603}, 125 (2001)
[arXiv:hep-ph/ 0103088].

\bibitem{GaLeFSE1}
J.~Gasser and H.~Leutwyler,
Phys.\ Lett.\ B {\bf 184} 83 (1987).

\bibitem{CH}
G.~Colangelo and C.~Haefeli, in preparation.

\bibitem{Marciano:2004uf}
W.~J.~Marciano,
arXiv:hep-ph/0402299.

\bibitem{MILC}
C.~T.~H.~Davies {\it et al.}  [HPQCD Coll.],
Phys.\ Rev.\ Lett.\  {\bf 92} (2004) 022001
[arXiv:hep-lat/0304004];\\
C.~Aubin {\it et al.}  [MILC Coll.],
arXiv:hep-lat/0309088.

\bibitem{UKQCD}
C.~R.~Allton {\it et al.}  [UKQCD Coll.],
arXiv:hep-lat/0403007.

\bibitem{qq+q}
F.~Farchioni, I.~Montvay and E.~Scholz  [qq+q Coll.],
arXiv:hep-lat/0403014.

\bibitem{Arndt:2004bg} 
D.~Arndt and C.~J.~D.~Lin, in Ref.~\cite{recent}.

\bibitem{heavy-light}
G.~Burdman and J.~F.~Donoghue,
Phys.\ Lett.\ B {\bf 280}, 287 (1992), \\
M.~B.~Wise,
Phys.\ Rev.\ D {\bf 45} (1992) 2188.\\
T.~M.~Yan {\it et al.},
Phys.\ Rev.\ D {\bf 46} (1992) 1148
[Erratum-ibid.\ D {\bf 55} (1997) 5851].


\end{thebibliography}
\end{document}